\documentclass{IEEEcsmag}

\usepackage[hyphens]{url}
\usepackage[colorlinks,urlcolor=blue,linkcolor=blue,citecolor=blue]{hyperref}
\hypersetup{breaklinks=true}
\usepackage{graphicx}
\usepackage{upmath}
\usepackage{doi}

\jvol{XX}
\jnum{XX}
\paper{X}
\jmonth{XX/YY}
\jname{Computing in Science and Engineering}
\pubyear{2024}
\begin{document}

 
\title{Scalable Delivery of Scalable Libraries and Tools: How ECP Delivered a Software Ecosystem for Exascale and Beyond}

\author{Michael~A. Heroux}
\affil{Sandia National Laboratories}

\markboth{CiSE Special Issue on Better Scientific Software: Improving while delivering for Exascale Computing}{}

\begin{abstract}
The Exascale Computing Project (ECP) was one of the largest open-source scientific software development projects ever.  It supported approximately 1,000 staff from US Department of Energy laboratories, and university and industry partners. About 250 staff contributed to 70 scientific libraries and tools to support applications on multiple exascale computing systems that were also under development.

Funded as a construction project, ECP adopted an earned-value management system, based on milestones. and a key performance parameter system based, in part, on integrations. With accelerated delivery schedules and significant project risk, we also emphasized software quality using community policies, automated testing, and continuous integration. Software Development Kit teams provided cross-team collaboration.  Products were delivered via E4S, a curated portfolio of libraries and tools.

In this paper, we discuss the organizational and management elements that enabled the efficient and effective delivery of ECP libraries and tools, lessons learned and next steps.

\end{abstract}

\maketitle
  
\begin{figure*}[h]
	\centering
	\includegraphics[scale=0.45]{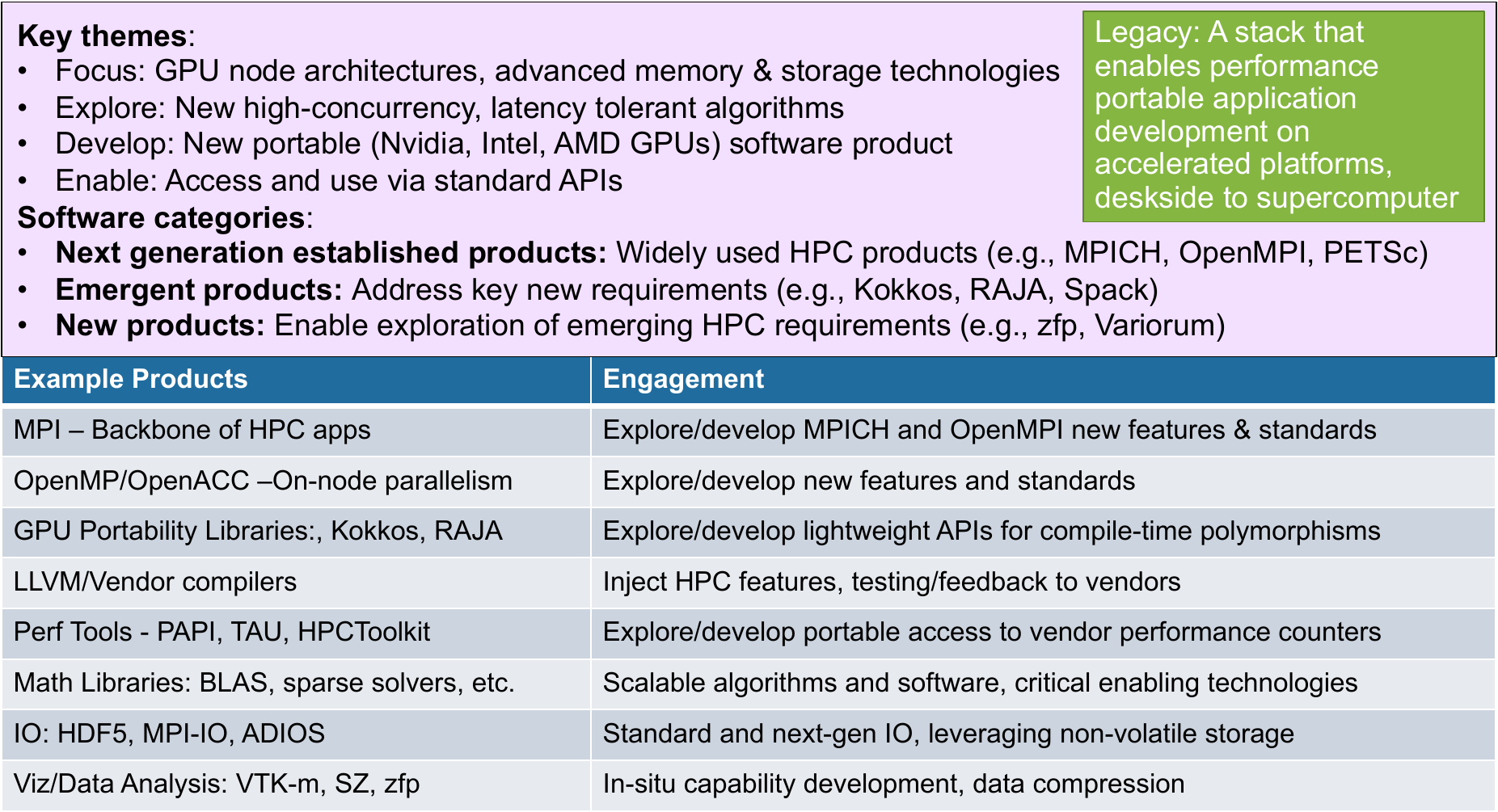} 
	\caption{\footnotesize ECP Software Technology efforts provided contributions to 70 scientific libraries and tools products.  Some products were well established and needed investments to provide portable and scalable execution on the target exascale systems.  Other products were still emerging at the beginning of ECP but were essential by the end.  Others were new at the start of ECP.  Two of these, zfp, a data compression library, and Variorum, a portable power management tool, received R\&D100 awards during the last year of ECP, a recognition of their future impact in high-performance computing.}
	\label{Products}
\end{figure*}

The Exascale Computing Project (ECP) represents one of the largest open-source scientific software development projects to date~\cite{ecp-kothe-lee-qualters-2019}\footnote{Exascale Website: \url{https://exascaleproject.org}}.  ECP funded approximately 1,000 scientists from the US Department of Energy (DOE) laboratory complex, and DOE university and industry partners.  The Software Technology Focus Area sponsored the efforts of about 250 people working on contributions to 70 open-source products. The result was a collection of reusable libraries and tools (see Figure~\ref{Products}) to support parallel applications and their portable execution on target platforms using GPU accelerators from three different computing system vendors. All of this work was done to support application codes that were still being designed and developed and for computer systems that were still being built.  

\subsection{ECP as a Construction Project}
Funded as a construction project, ECP used a tailored earned-value management (EVM)~\cite{evm} system.  Flexibly using EVM to structure work activities, we were able to design, develop, and deliver our software contributions across a wide range of teams, products, and organizations. Use of EVM enabled us to effectively and efficiently manage the efforts of our 35 teams, and detect potential staffing issues, technical challenges and misaligned efforts in need of correction or offramping.  These efforts were successful in the formal sense of passing the key performance parameters (KPPs) defined for the project.  Even more important, the work received mention in the \textit{CHIPS and Science Act (page 175)}~\cite{chips-and-science-act-2022} which specifically called out the need to sustain the ECP software ecosystem we developed.

In this paper, we describe the challenges ECP represented and the organizational and management approaches that enabled the development and delivery of ECP libraries and tools to the open-source software community.  We also discuss the lessons learned from this effort and some future directions.

\subsection{ECP Challenges}
ECP was a large, multi-institutional, multi-year project to design and develop new application, library, and tool capabilities on a new generation of systems designed to execute a billion-billion ($10^{18}$ or exa) operations per second.  The key exascale systems are named Frontier at Oak Ridge National Laboratory, Aurora at Argonne National Laboratory, and El Capitan at Lawrence Livermore National Laboratory.  Frontier and El Capitan use AMD GPUs, and Aurora uses Intel GPUs.  When ECP started, system details were still evolving and Aurora was based on a different (non-GPU) processor.  Even so, we knew these systems would require disruptive changes to scientific problem formulations, algorithms, and software design.  Most of the performance from these systems would come from the GPUs, capable of trillions ($10^{12}$) of operations per second. GPUs had been used in previous systems but not with the diversity of architecture or the scale of performance needed to reach the exascale threshold, nor with the requirement for performance portability across multiple kinds of GPUs. The challenge for the ECP leadership team was to provide a framework for successful software development in this environment.

\section{Organizational Elements}

\begin{figure*}[h]
	\centering
	\includegraphics[scale=0.45]{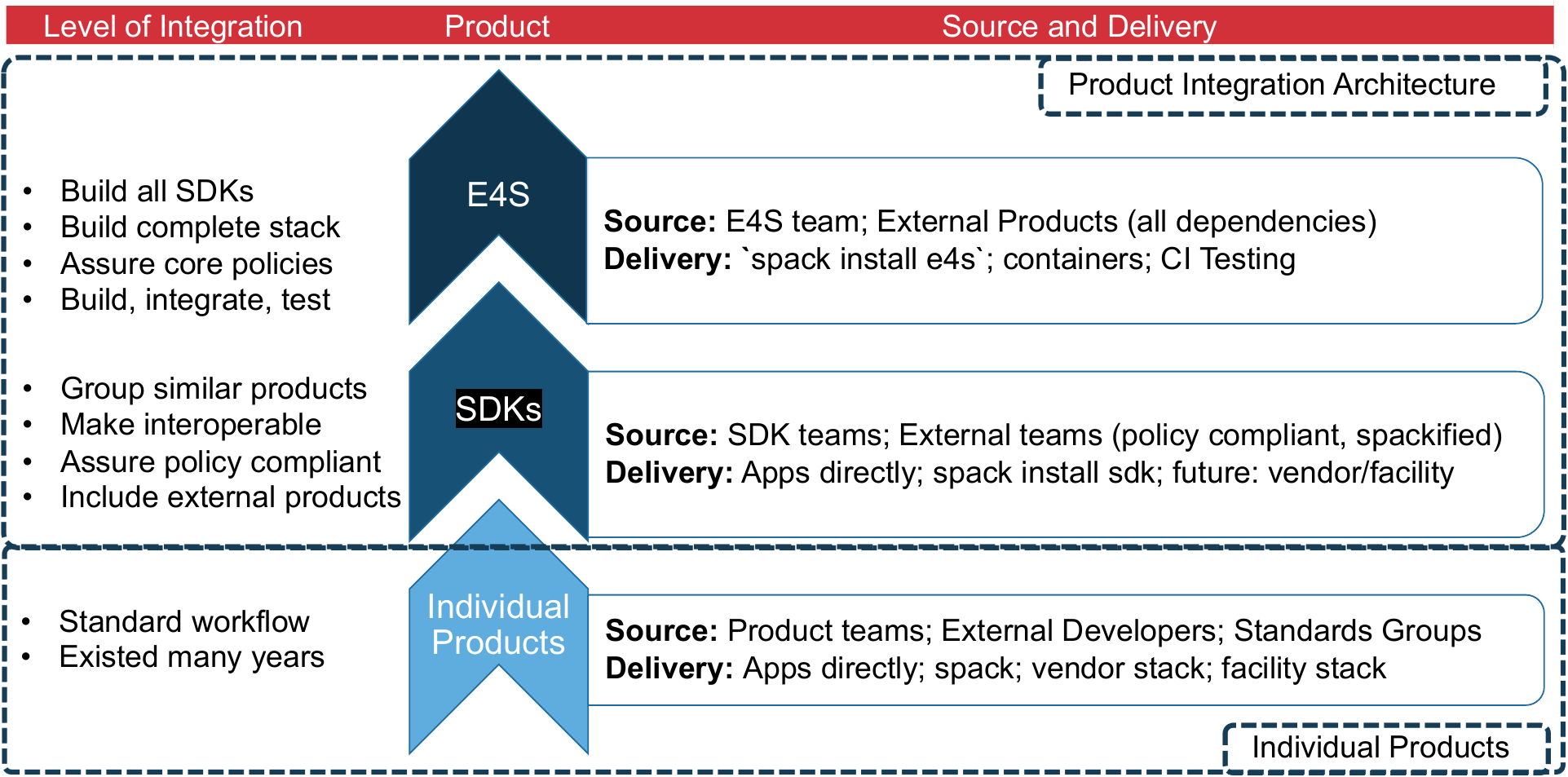}
	\caption{\footnotesize ECP software libraries and tools efforts were organized at three levels: individual products, Software Development Kits (SDKs), and the Extreme-scale Scientific Software Stack (E4S).  The product teams were responsible for defining and executing their milestones and integrations using their own internal processes, as long as the teams achieved their milestones and integrations.  SDK teams were responsible for coordinating activities among product teams.  The E4S team was responsible for coordinating activities among SDK teams and curating the full collection of ECP libraries and tools.  ECP leaders were responsible for managing the entire process.}
	\label{Product-SDK-E4S}
\end{figure*}

One key element of ECP library and tools efforts was its organizational structure, designed to provide a framework for teams to collaborate and succeed in a challenging environment with flexibility to accommodate team and project needs.  The structure had three layers: product teams, Software Development Kit (SDK) teams, and the Extreme-scale Scientific Software Stack (E4S) team (Figure~\ref{Product-SDK-E4S}).

\subsection{Product Teams}
The fundamental ECP organizational building block was the product team.  Milestones and integrations (see the next section on management) were defined and tracked per product.  Product teams were responsible for defining and executing their activities and integrations and managing their own internal processes.  Product plans were reviewed by ECP leadership to ensure feasibility and alignment with ECP end-project objectives.  Product teams have always been the fundamental building block of DOE software development efforts.  However, within ECP their work was coordinated and supported by the SDK and E4S teams, and ECP leadership provided a framework for planning, executing, tracking, and assessing progress. The result was a structured, but flexible, approach to software product development.

\subsection{SDK Teams}
The next building block was the Software Development Kit (SDK) team.  Each SDK team was a collection of product teams producing compatible and complementary products.  The purpose of SDKs was to facilitate interaction among product teams in a variety of activities such as requirements, design space exploration, training, and the evolution of software practices and tools. Coordination of versioning, vendor interactions, design space exploration, and software delivery at the SDK level helped amortize costs across related products and reduced complexity at the top software stack level.  This level also spurred the dynamics of \textit{co-opetition}~\cite{coopetition} where teams collaborated on some activities and learned from each other.  Then, to make sure they kept pace with other SDK members, teams competed to be the best in other activities.  This dynamic was very effective at accelerating progress and improving quality.

\subsection{E4S Team}
The final building block was the E4S\footnote{E4S website: \url{https://e4s.io}} team, which focused on building, testing, delivering, and deploying the complete software stack. E4S (the product) is a curated build of more than 100 primary products and their dependency (hundreds more products). Using Spack, E4S incorporates external, including proprietary, products.  E4S can also be configured to create custom builds that target many environments---from laptops to clusters, leadership systems, and edge computing environments. E4S comes in containerized environments and has binary caches of previously built products to enable rapid rebuilds.  The E4S team manages versioning across products for reproducibility, correctness, and security patching.  Actively managing a full portfolio of configurable HPC software products that are built and tested on a variety of HPC platforms, including on the leadership platforms under development, provides a rich resource of version-compatible products, built under numerous different parameter settings, and tested on many systems.

\subsection{Other Delivery Mechanisms}
While many ECP libraries and tools products were (and still are) delivered through the SDKs and E4S, most products were also delivered directly to clients as ``second-party''\footnote{Second-party software is a distinct product developed in close collaboration with another product.} software that was co-developed with their application users.  Finally, some of our work was delivered to community ecosystems such as LLVM\footnote{LLVM: \url{https://llvm.org}}, a major platform for compiler and runtime library features and to community efforts like the C++ Language standard.

\section{Efficient and Effective Management} \label{management}
\begin{figure*}[h]
	
	\centering
	\includegraphics[scale=0.45]{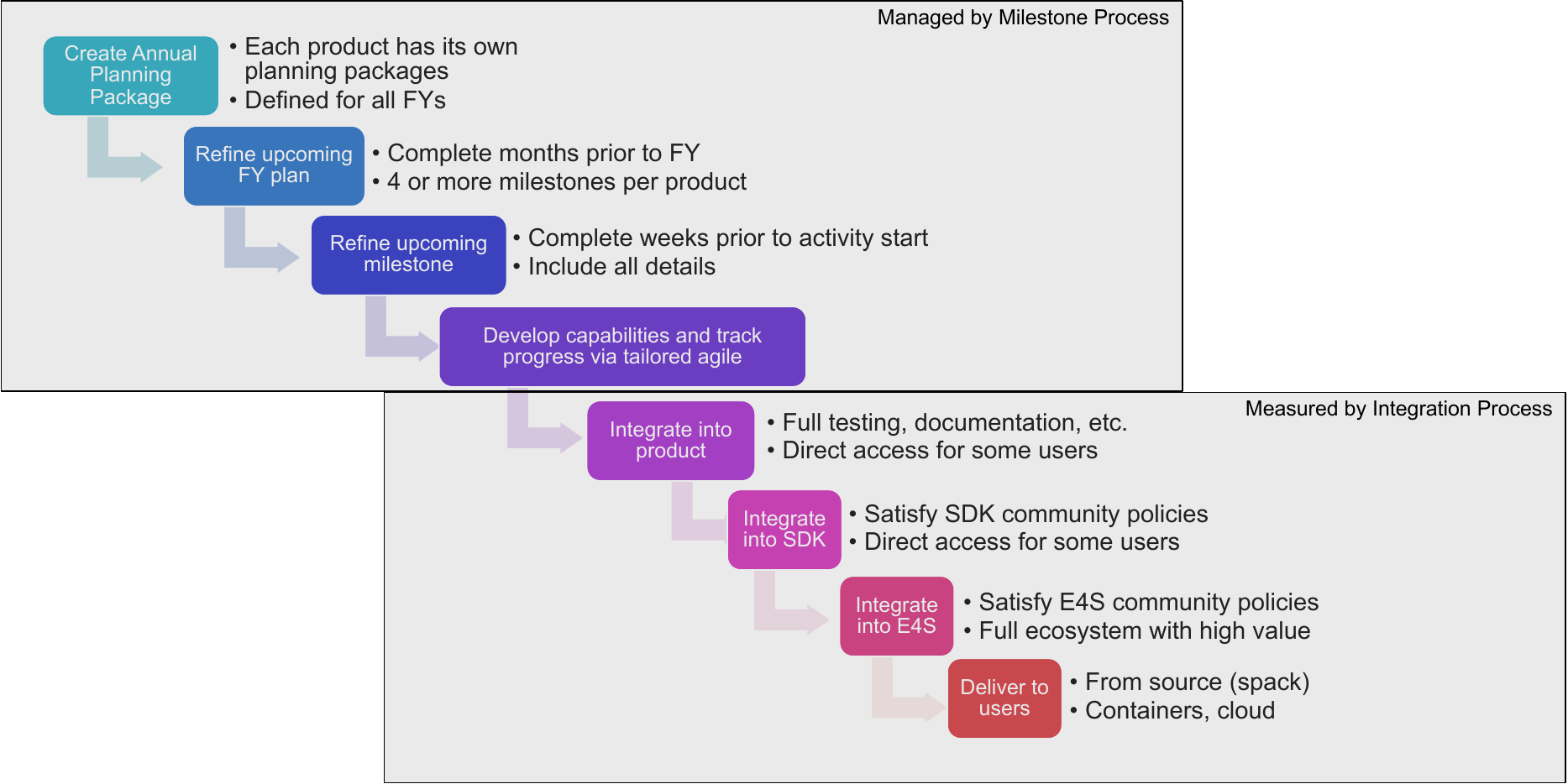}
	\caption{\footnotesize ECP library and tool development and delivery process: The above diagram shows the idealized workflow for ECP library and tool development and delivery.  The process began with a high-level objective described in per-product, per-year planning packages (short paragraphs describing what will be done) to provide libraries and tools for ECP applications to use on exascale systems by the end of the project.  Each year, planning packages were refined to describe specific activities as a collection of milestones.  As capabilities were developed, they were integrated into the ecosystem and demonstrated in the client environment. The process was repeated until the project was completed.}
	\label{Milestone-Integration}
\end{figure*}

Trusted systems are often the result of addressing two complementary concerns: doing things right and doing the right things.  The first concern is about quality, and the second is about value.  For ECP efforts, we focused on milestones and integrations. Milestones incentivized and supported teams in planning and execution toward ``doing things right'' and integrations incentivized and measured efforts toward ``doing the right things.''  In this section, we describe how we used milestones and integrations to plan, execute, track, assess, and adapt ECP software libraries and tools efforts. Figure~\ref{Milestone-Integration} shows the relationship between milestones and integrations and how they were used to support the ECP software lifecycle.

\subsection{Milestones}
An idealized project mental model has three main components: creating something (scope), over a certain timespan (schedule), for a certain amount of effort and resources (cost).  ECP was executed using EVM, an approach that packetizes scope, schedule, and cost as a collection of \textit{activities}. In ECP, each activity was defined for a particular product over a particular time (usually a few months) containing a description of what would be done (its scope), the amount of budget for the activity, and the expected beginning and end dates. On a large project like ECP, many activities can be executed concurrently, and the set of in-progress activities represents a wavefront in time, as current activities are completed and new activities are started. ECP's orchestrated and concurrent execution of many activities using EVM enabled the scalable delivery of many products by many teams in an orderly manner.

If all activities were perfectly predictable, EVM would result in a smooth progression of completing milestones (the successful end of an activity) resulting in an on-time delivery of all capabilities for the estimated cost. However, most projects are not so predictable, especially software R\&D projects like ECP that are producing new capabilities to support new applications for new computers that themselves are being co-designed and developed along with the software.  The ability to effectively monitor and adapt activities was what made EVM so valuable to ECP.

EVM has many project management and control concepts.  The two that were most useful for ECP product development were \textit{cost performance index (CPI)} and \textit{schedule performance index (SPI)}:
\begin{itemize}
	\item \textbf{CPI:} Measures actual costs vs. predicted costs.  A CPI of 1.0 means actual costs match predicted costs at a given point in time of the project.  A CPI above (below) 1.0 means costs are less (more) than predicted.
	\item \textbf{SPI:} Measures actual progress vs. predicted progress in completing work (scope).  An SPI of 1.0 means work is on schedule.  An SPI above (below) 1.0 means work is ahead (behind) schedule.
\end{itemize}

In ECP, we used many EVM metrics but CPI and SPI  were particularly helpful in getting an early indication of when a particular library or tool product team was struggling.  For most projects, their CPI and SPI were very close to 1.0 throughout the project.  This did not necessarily mean that ECP teams were perfect at estimating.  For CPI, it typically meant that team staffing was stable and the team was spending money at a consistent rate to pay salaries. For SPI, it meant that features were being delivered on time.  However, because features were not always described in great detail, the features delivered could vary and still satisfy what was planned.  In particular, an elaborately planned feature could be trimmed and still be declared as completed.

While most teams consistently had CPI and SPI values near 1.0, some teams had CPI, SPI, or both, values significantly less than 1.0.  Usually, these teams needed help to refine scope, increase funding, relax schedules, or make some other adjustment, to make it possible for them to succeed.  The ECP leadership team was able to identify these concerns early and work with teams to make adjustments.  In other instances, poor CPI and SPI values signaled deeper problems such as the loss of key staff or poor team morale, problems that were harder to address but important to track.  Monthly reviews of CPI and SPI were key elements of the ECP project management approach. Tracking SPI and CPI was used constructively (not punitively) to help teams succeed and keep track of progress at all levels of the project.

The end objective of ECP for the scientific libraries and tools teams was very clear: Produce a software stack that would support applications on the target exascale systems before the end of the project.  However, the exact path to that objective was not known at the beginning of the project and was disrupted regularly as the project proceeded. We had changes in the target systems, application requirements, development team staffing, and external risk factors that arose or diminished as the project proceeded. 

Because of the certainty of our objective and the uncertainty about our path toward that objective, the ECP activity (milestone) planning process contained end-goal planning objectives and coarse grain plan descriptions that were refined as the project proceeded.  Each year, we would define approximately 300 activities and at any given time more than 100 activities would be in progress.

\begin{figure*}[h]
	
	\centering
	\includegraphics[scale=0.45]{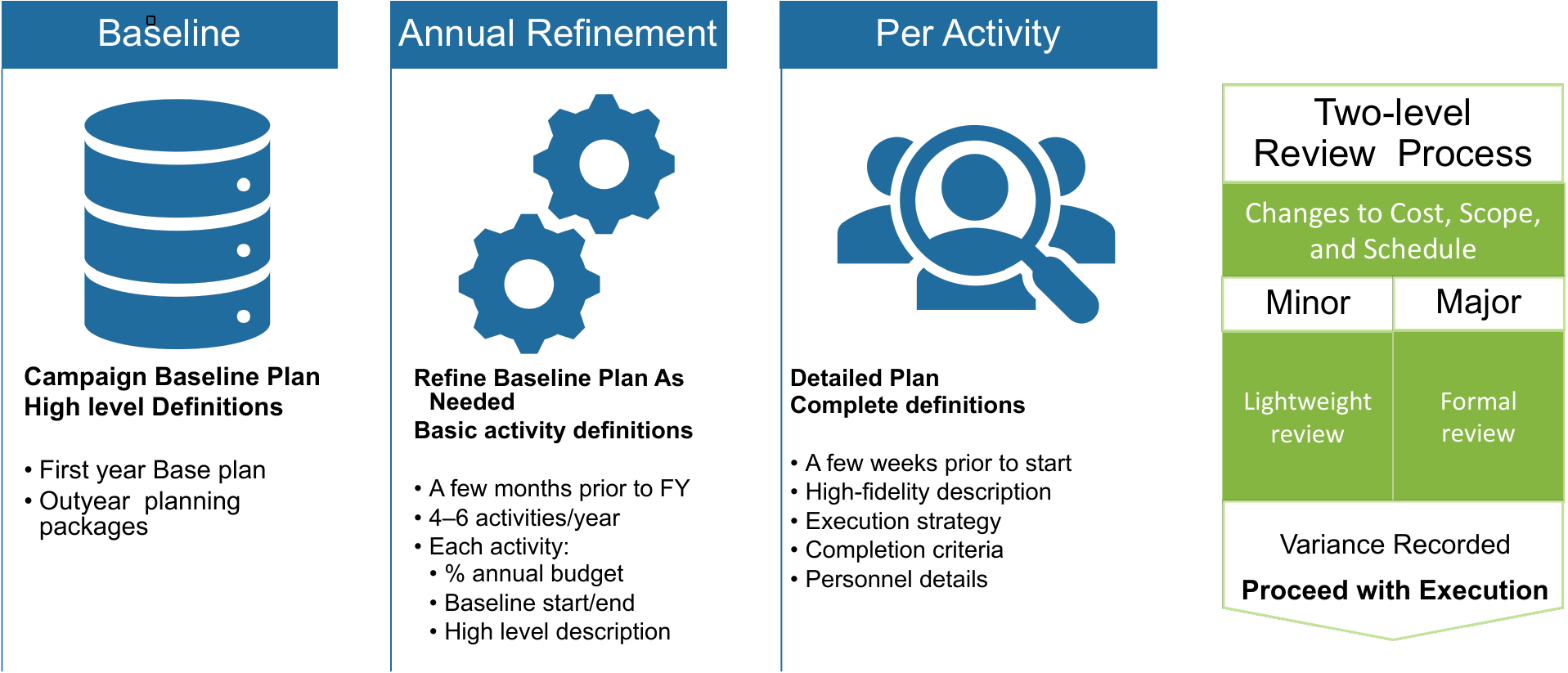}
	\caption{\footnotesize ECP libraries and tools activity/milestone planning process with annual refinement and change management: The above diagram shows the planning process for ECP library and tool development and delivery.  The process started with a multi-year, coarse-grain plan for the entire project, for each product. A few months before the start of a new fiscal year, each product team would refine that year's planning package.  A few weeks before the start of an activity, the remaining details were added. The process was repeated until the project was completed.  For any mid-year changes, a change management process was used to ensure that the changes were well understood, agreed to by all stakeholders, and made visible to all staff.}
	\label{Annual-Campaign}
\end{figure*}

As shown in Figure~\ref{Annual-Campaign}, the process began with a multi-year baseline plan that provided coarse-grain descriptive paragraphs (called planning packages), one planning package per product per year, for each year of the project. The final objective for each product was a description of the expected final product capabilities for the exascale systems.  A few months before the start of a new fiscal year, each product team would refine that year's planning package into 4 - 6 activities for the year with an estimate of the percent annual budget for each activity, a baseline start/end date and a high-level description.  A few weeks before the start of an activity, the remaining details needed for staffing, completion criteria, etc., were added.  Outyear planning packages could also be updated.  The process was repeated until ECP was completed.

To manage changes in scope, schedule, or costs that were needed outside of the annual planning process, we had a two-level change management process.  The key objective of the process was to make sure we clearly articulated proposed changes, reviewed changes with transparency (especially with sponsors and lab management), and then recorded our decisions for later reference.  The process was not intended to prevent changes but to ensure that changes were agreed to by project stakeholders and well understood by everyone involved in the project to retain transparency and trust.  We consistently advised product teams that they should always be doing the most important work and, if what was most important changed, then they needed to change their plans to support the change in priorities.

\subsection{Integrations}
\begin{figure*}[h]
	
	\centering
	\includegraphics[scale=0.45]{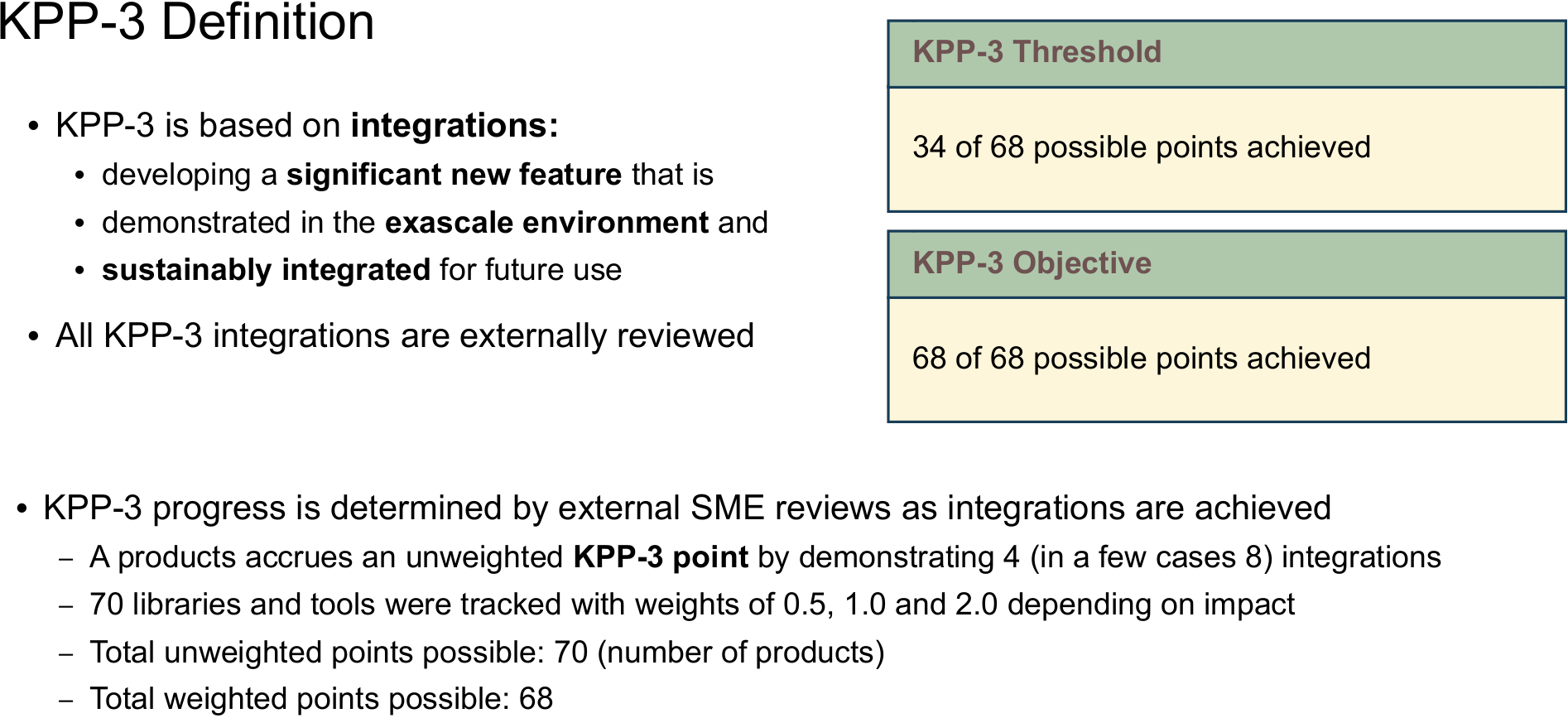}
	\caption{\footnotesize The definition of KPP-3 was centered on the \textit{sustainable integration of significant capabilities}, called an \textit{integration}.  Each product team had to demonstrate four (or in some cases eight) integrations during the lifetime of the project.  Integrations were reviewed by external subject-matter experts (SMEs).  Artifacts such as screenshots of output and client letters were used as evidence of integrations and SME reports were used to determine if the KPP-3 requirements were satisfied. Final approval was given by the ECP Federal Project Director, who had formal approval authority.}
	\label{KPP-3-Definition}
\end{figure*}

ECP had four key performance parameters (KPPs) which were metrics used to assess overall project success. Two KPPs were focused on applications, another on hardware, and one on software libraries and tools.  The software KPP, called KPP-3, is sketched in Figure~\ref{KPP-3-Definition}. The KPP-3 definition was intended to be accurate at a high level and flexible, to ensure that the software developed by ECP teams was integrated into client environments in a meaningful and sustainable manner while allowing for a variety of integration approaches.

Activities and milestones were how ECP addressed the ``doing things right'' goal.  Integrations were how ECP addressed the ``doing the right things'' goal.  Integrations were the process of taking the features developed by the teams and sustainably integrating them into client environments.  

The primary emphasis of KPP-3 was to ensure a ``capable and sustainable software ecosystem,'' a long-term view, where software capabilities are not just developed but are maintained and usable over time. The integration goal prioritized real-world applicability and long-term use, starting with pre-exascale environments and then transitioning to exascale environments. The integration goals were product-focused and success in integration was quantified by counting the number of ``integrations,'' the number of successful capability integrations of a product into client environments providing a concrete metric to evaluate doing the right things.

\subsection{Annual Lifecycle}
Using the organizational elements of milestones and integrations, ECP implemented an annual planning, execution, tracking, reporting, assessing, and adapting lifecycle.  The lifecycle centered on the product teams and their milestones and integrations.  
\begin{itemize}
	\item \textbf{Planning:}  The lifecycle began with the activities planning process.  Each product team refined its activities and resulting milestones annually, adding final details just before starting the execution of an activity.  The SDK and E4S teams did the same. The planning process was reviewed by the ECP project leadership team to help ensure rigorous and realistic plans that were consistent across products and in line with the long-term ECP objective of a capable and sustainable ecosystem.
	\item \textbf{Execution:}  Execution of activities toward milestone completion progressed during the year.  Teams generally worked independently, engaging with their clients as needed.  The SDK and E4S teams engaged with product teams to support their efforts and to coordinate activities among teams.
	\item \textbf{Tracking:}  Progress was tracked using Jira and Confluence, two widely used, web-based project management platforms.  ECP used custom Jira issue types that aligned with and could automatically synchronize with a tool called Primavera\footnote{Use of Primavera was a requirement of our ECP sponsors but development teams never had to work with it directly because we mirrored and synced content between Primavera and Jira.}.  Monthly reporting of CPI and SPI helped identify potential problems early.
	\item \textbf{Reporting:} Concurrent with the assessment process was the production of a Capability Assessment Report (CAR)~\cite{ECP_CAR3.0}.  The CAR provided a detailed assessment of the ECP Software Technology efforts including a description of activities, progress, and plans for each product. The CAR was used to inform the annual project reviews and identify gaps in the project. It also provided stakeholders and the scientific community with project updates.
	\item \textbf{Assessing:}  Annual project reviews were conducted with project leadership and external subject-matter experts (SMEs).  The SMEs reviewed the progress of the project and provided feedback to the project leadership team.  The project leadership team used the SME feedback to make adjustments to the project.  
	\item \textbf{Adapt and Repeat:}  The annual review process led to numerous project changes over the years of ECP, including restructuring teams, off-ramping products, and addressing product gaps using contingency funds\footnote{By monitoring milestone and integration progress, the ECP leadership team was able to see when products were not progressing (as evidenced by repeatedly missing milestones), or had little probability of making an impact, as evidenced by a lack of pre-exascale integrations, or both.  In some cases, corrections in planning and strategy led to improvements.  In other cases, we created an exit plan to stop funding the project.  Similarly, sometimes we identified gaps in our portfolio and initiated new activities for new products to address the gaps.}. The lifecycle was repeated annually until the project was completed.
\end{itemize}

\subsection{Complementing Product Efforts}
The ECP approach to the development and delivery of capabilities through a milestone and integration process was designed to flexibly complement the software engineering practices of individual product teams.  While product teams were expected to address ECP community policies\footnote{E4S Policies: \url{https://e4s-project.github.io/policies.html}} for software quality, user support, and integration, the ECP leadership team did not direct product teams to use specific software engineering practices, tools, or processes beyond the use of Jira and Confluence for ECP-specific efforts. 

ECP was a project with a fixed timeline. Most product teams existed before ECP and continued their efforts after ECP finished, and many received concurrent funding from non-ECP sources.  By requesting only high-level, coarse-grain information from project teams, we minimized duplicate record-keeping.  ECP's use of milestones and integrations addressing the specific funding scope of ECP could easily integrate with the finer-grain workflows of a product team.  Also, because we used the web-based tool Jira for ECP issue tracking, cross-referencing ECP issues with other systems, e.g., GitHub Issues, was easily done through web links.  

\section{Lessons Learned}

ECP was an ambitious project that was successful in many ways and yet still had signficant room for improvement.  It was successful in the formal sense of passing the key performance parameters (KPPs) defined for the project.  We learned many lessons that will be useful for future projects.  The following are some of the most important:

\begin{enumerate}
	\item \textbf{Organizing independently-developed products into a coherent software ecosystem provides value:}  ECP delivered its milestones, integrations, and a full stack of portability GPUs libraries and tools to support ECP applications on exascale systems.  Success is evident from the explicit support for sustaining this ecosystem after ECP ended.
	\item \textbf{Organizing a research software project using a tailored EVM structure can work:} Using EVM with a two-level refinement planning process enabled keeping the long-term project objective (a capable software stack for exascale systems) in focus and permitted adaptation due to uncertainty and unforeseen changes in external factors.  The EVM structure also provided a mechanism for identifying teams that were struggling and needed help.
	\item \textbf{The use of milestones and integrations resulted in efficient and effective efforts:}  Milestones supported doing things right. Integrations supported doing the right things. 
	\item \textbf{Heavyweight processes were expensive to support:}  ECP was a large and visible project with very specific reporting requirements. Most requirements were handled by ECP leadership but KPP content required individual technical staff members to produce evidence and formal reports for SMEs.  This process was necessary for assuring our federal project director that we were meeting our KPP-3 requirements but was costly.  In the future, we will look for ways to reduce the overhead of producing evidence and formal reports.
\end{enumerate}

\section{Next Steps}
The Exascale Computing Project finished its development and delivery of libraries and tools in December 2023.  However, the end of ECP is the beginning of the ``Exascale Era.'' The following are some of the next objectives for the software ecosystem ECP started:

\begin{itemize}
\item \textbf{Achieving 100X performance and efficiency:} ECP demonstrated 100X or more increase in performance and energy efficiency~\cite{heroux2023nasa}.  However, ECP impacted only tens of applications and software technologies directly and focused only on very high-end systems.  Hundreds of scientific software products can realize this ``100X'' potential, so the work continues. We need to make the ECP libraries and tools available for high-end systems, deskside and rack systems, and computing centers with limited power budgets where energy efficiency is key.  Many public and private sector scientific software stacks are primarily running on CPU-based parallel computers, limiting performance and energy efficiency improvements.  Bridging performance and efficiency gaps for these communities is essential.

\item \textbf{Delivering a turnkey community scientific software stack:}  The ECP project has been delivering a near-turnkey scientific software stack to the HPC/AI community for the past three years.  Spack-enabled software stacks, with E4S as a primary deliverable, have been key components of the ECP software delivery strategy. Part of our vision is making this stack available ubiquitously---critical as we build toward the emerging area of HPC/AI for science---incorporating and complementing vendor-provided libraries and tools.  

\item \textbf{Realizing the cross-cutting potential of AI/ML for science:} AI/ML workflows for scientific discovery will be physics-informed machine learning, graph neural networks, deep reinforcement learning, and multimodal large language models, to enable digital twins, analysis of scientific data for discovery and acceleration of traditional simulations.  Presently, many scientists manage their own AI/ML software stacks.  The ECP software ecosystem can be readily augmented beyond its existing support of mainstream AI/ML libraries (e.g., TensorFlow, PyTorch) to include new libraries and tools that support AI for science.

\item \textbf{Sustaining a collaborative approach for scientific software development:}  ECP demonstrated successful large-scale scientific software development delivering a curated software stack to the HPC/AI community.  The ECP leadership approach has been successful in providing a framework for planning, executing, tracking, assessing, and reporting on the project.  We will sustain these efforts and expand them to other communities.
\end{itemize}

\section{Conclusions}
The Exascale Computing Project provided a unique opportunity for the DOE scientific libraries and tools communities by sponsoring sustained multi-year funding that promoted cross-institutional collaboration at a scale the DOE high-performance computing community had not experienced before. 

Using a three-tier organizational structure (product, SDK, E4S), ECP enabled mostly autonomous activities for product teams, rapid design space exploration via SDKs, and delivery of a curated software stack via E4S.  Using a tailored EVM system ECP was able to execute hundreds of activities in parallel and still carefully monitor progress.  The use of EVM framed the expected scope, schedule, and cost, providing a foundation for expected outcomes and allowing us to focus our attention on monitoring and addressing variations and realize the scalable delivery of our libraries and tools.  The outcome is a GPU-capable, performance-portable stack of 70 libraries and tools, representing 1,700 completed milestones and nearly 300 documented integrations over six years.

The legacy of these efforts is proof of the potential of portable accelerated computing, project leadership structures and strategies for realizing success, and a collection of libraries and tools that, in combination, can deliver two orders of magnitude performance and energy efficiency improvements for scientific applications, the result of the efforts of hundreds of scientists who were part of ECP.  We look forward to building on this legacy in the years to come.

\section{Acknowledgment}
We thank the many ECP staff members who contributed to the success of the ECP software libraries and tools efforts. Their investments of time and energy made the plans and processes work despite challenges and initial uncertainties.  There would be nothing to write about without their efforts.
This research was supported by the Exascale Computing Project (17-SC-20-SC), a collaborative effort of the U.S. Department of Energy, Office of Science, and the National Nuclear Security Administration. Sandia National Laboratories is a multimission laboratory managed and operated by National Technology and Engineering Solutions of Sandia, LLC., a wholly owned subsidiary of Honeywell International, Inc., for the U.S. Department of Energy's National Nuclear Security Administration under contract DE-NA0003525.

\bibliographystyle{plain}
\bibliography{CiSE-ScalableDelivery-Heroux}

\begin{thebibliography}{1}

\bibitem{coopetition}
Adam Brandenburger and Barry Nalebuff.
\newblock The rules of co-opetition.
\newblock {\em Harvard Business Review}, January 2021.
\newblock Accessed: 2021-11-02.

\bibitem{chips-and-science-act-2022}
{CHIPS and Science Act}.
\newblock \url{https://science.house.gov/chipsandscienceact} and
  \url{https://science.house.gov/imo/media/doc/the_chips_and_science_act.pdf}.
\newblock United States Congress, July 2022.

\bibitem{heroux2023nasa}
Michael~A. Heroux.
\newblock {100X}: Leveraging the future potential of {US} {Exascale Computing
  Project} investments.
\newblock \url{https://www.nas.nasa.gov/pubs/ams/2023/06-20-23.html}, 6 2023.
\newblock Presentation at the NASA Advanced Modeling \& Simulation Seminar
  Series.

\bibitem{ECP_CAR3.0}
Michael~A. Heroux, Lois~Curfman McInnes, Rajeev Thakur, Jeffrey~S. Vetter,
  Xiaoye~Sherry Li, James Ahrens, Todd Munson, Kathryn Mohror, Terece~L.
  Turton, and {ECP Software Technology teams}.
\newblock {ECP Software Technology Capability Assessment Report, V3.0}, June
  2022.
\newblock \url{https://www.osti.gov/biblio/1888898}.

\bibitem{ecp-kothe-lee-qualters-2019}
D.~Kothe, S.~Lee, and I.~Qualters.
\newblock Exascale computing in the {United States}.
\newblock {\em Computing in Science and Engineering}, 21(1):17--29, 2019.
\newblock \doi{10.1109/MCSE.2018.2875366}.

\bibitem{evm}
{U.S. Department of Energy, Office of Project Management}.
\newblock {Earned} {Value} {M}anagement.
\newblock
  \url{https://www.energy.gov/projectmanagement/earned-value-management}.
\newblock Accessed: 2023-10-01.

\end{thebibliography}

\begin{IEEEbiography}{Michael A.~Heroux} is a senior scientist at Sandia National Laboratories and Scientist in Residence at St. John's University, MN. His research interests include all human and technical aspects of scalable scientific and engineering software for new and emerging parallel computing architectures.\end{IEEEbiography}

\end{document}